\documentclass[twocol]{ametsocV6.1}
\usepackage{rotating}

\title{A multi-frequency spaceborne radar perspective of deep convection}

\authors{Randy J. Chase\aff{a,b}\correspondingauthor{Randy J. Chase, randy.chase@colostate.edu}, Brenda Dolan\aff{b}, Kristen L. Rasmussen\aff{b}, Richard M. Schulte\aff{b}, Graeme Stephens\aff{b,c}, F. Joe Turk\aff{c}, Susan C. van den Heever\aff{b}}

\affiliation{
\aff{a}{Cooperative Institute for Research in the Atmosphere, Colorado State University, Fort Collins, CO} \\
\aff{b}{Department of Atmospheric Science, Colorado State University, Fort Collins, CO, USA} \\
\aff{c}{Jet Propulsion Laboratory, California Institute of Technology, Pasadena, CA 91109, USA} \\
}

\abstract{Global numerical weather models are starting to resolve atmospheric moist convection which comes with a critical need for observational constraints. One avenue for such constraints is spaceborne radar which tend to operate at three wavelengths, Ku-, Ka- and W-band. Many studies of deep convection in the past have primarily leveraged Ku-band because it is less affected by attenuation and multiple scattering. However, future spaceborne radar missions might not contain a Ku-band radar and thus considering the view of convection from Ka-band or W-band compared to the Ku-band would be useful. This study examines a coincident dataset between the Global Precipitation Measurement (GPM) Mission and CloudSat as well as the entire GPM record to compare convective characteristics across various wavelengths within deep convection. We find that W-band reflectivity (Z) tends to maximize near the Ku-band defined echo-top while Ka-band often maximizes 4-5 km below. The height of the maximum Z above the melting level for W-band does not linearly relate to the Ku-band maximum. However, using the full GPM record the Ka-band 30 dBZ echo-tops can be linearly related to the Ku-band 40 dBZ echo-top with an $R^2$ of 0.62 and a root mean squared error of about 1 km. The spatial distribution of echo-tops from Ka-band corresponds well to the Ku-band echo-tops, highlighting regions of relatively large ice water path. This paper suggests that Ka-band only missions, like NASA's Investigation for Convective Updrafts, should be able to characterize global convection in a similar manner to a Ku-band system.}

\begin{document}

\maketitle

%
%
%
\statement

There has been a long history of studying global storms using Ku-band [2 cm, 13 GHz] spaceborne radar, most likely because of the least number of challenges (e.g., loss of signal) at Ku-band compared to Ka- [8 mm, 35 GHz] and W-band [3 mm, 89 GHz]. However, each radar system offers different perspectives on hydrometeor profiles observed in deep convection that have remained largely unexplored, perspectives that provide insights into convective storm systems. Therefore, it is useful to know how storms measured at Ka-band and W-band compare to storms measured at Ku-band. We find that many of Ka-band convective properties can be linearly related to Ku-band and thus Ka-band only mission designs should be suitable for studying convective storms. 
%
%
%

%

\section{Introduction}

For many places around the world, direct observations of storm characteristics are difficult to obtain. Regions over oceans, complex terrain, and other remote areas without access to high temporal and spatial resolution ground-based remote sensing or in-situ observations are common across the world. Evaluating the representation of storms in global numerical weather predictions is therefore a major challenge. This is now becoming particularly problematic since global numerical weather prediction models have grid-spacings that are approaching convective allowing scales \citep[e.g., 4 km: DYAMOND;][]{Stevens2019}. One solution to help observe storms globally is to use spaceborne cloud and precipitation radars \citep[see ][for a review]{Battaglia2020b}.

The Tropical Rainfall Measurement Mission \citep[TRMM;][]{Kummerow1998} flew the first spaceborne precipitation radar that was launched in 1997 and operated at 13.5 GHz (i.e., Ku-band; 2 cm wavelength). Its focus was on estimating near-surface rainfall around the tropics. The first spaceborne cloud radar was launched in 2006, as part of the CloudSat mission. The CloudSat Cloud Profiling Radar \citep[CPR; CloudSat ][]{Stephens2002} was a highly sensitive (-30 dBZ) 94 GHz (W-band; 3 mm wavelength) cloud radar with a focus on studying cloud systems. Following CPR and TRMM was the Global Precipitation Measurement (GPM) mission \citep{Hou2014} that was launched in 2014 and included the Dual-frequency Precipitation Radar (DPR), operating at 13 GHz (KuPR) and 35 GHz (KaPR) on the GPM core satellite. GPM aimed to continue the efforts of TRMM. Following GPM, NASA launched a Radar in a Cubesat \citep[RainCube; ][]{Peral2018} technology demonstration of a Ka-band precipitation radar housed on a small spacecraft (Cubesat), which provided evidence that scientific radar data could be collected by smaller-scale missions at lower costs than its predecessors (e.g., TRMM, GPM). This point was further underscored in 2023 by the private company tomorrow.io who launched two \textit{pathfinder} Ka-band radars built based on RainCube heritage with plans of launching a fleet of Ka-band scanning radars and microwave radiometers in the next few years \citep{Roy2023}. After RainCube, a spaceborne radar was launched by China in 2023, named the FY-3G, carrying a nearly identical radar to GPM \citep{Zhang2023}. The most recent spaceborne radar, ESA’s EarthCARE mission, was launched in May 2024, carrying a near-copy of the CloudSat CPR \citep{Illingworth2015}. 

Apart from EarthCARE, none of the previous spaceborne radars have Doppler capabilities, and except for CPR, are generally focused on measurement of precipitation size hydrometeors near the surface. To date, the current spaceborne radar record does not have any direct dynamic information to characterize the dynamical processes defining the intensity of storms which would be valuable for assessing the representation of atmospheric moist convection in Earth system models. Additionally, the potential value of Doppler radar data from space (e.g., EarthCARE) for this purpose remains an unresolved topic of research, but has been investigated using radar forward simulators and numerical weather prediction \citep{Kollias2022}. As an alternative, convective proxies derived from radar reflectivity profiles, like a storm's height at a given reflectivity threshold, have been derived. The general idea for reflectivity-only convective metrics is that the higher the echo-top height of a specific reflectivity value, the more intense the convection. The reasoning is that a stronger updraft can loft more massive hydrometeors to higher altitudes, thus producing larger reflectivities at higher altitudes. The exact reflectivity value adopted for the echo-top calculation varies because of the different saturation values for each wavelength and the radars’ minimum sensitivity thus offering different insights about the lofting of hydrometeors on convection. Reflectivity values for convective proxies range from 0 and 10 dBZ for CPR \citep[e.g.,][]{StephensandWood2007, Luo2008, TakahashiandLuo2014} and 30 to 40 dBZ for TRMM \citep[e.g.,][]{Zipser2006,Houze2015} and GPM's KuPR \citep[e.g.,][]{Skofronick-Jackson2018}. The broad result of the echo-top convective proxies between missions paints a consistent picture of stronger convection being generally located over land, with more frequent locations of intense convection over hotspots like Central Africa, North-Central Argentina and the central United States \citep{Zipser2006,Houze2015,TakahashiandLuo2014}. The differences in echo-top heights of the different radar systems, not yet researched in detail, also offers potential insights into processes associated with detraining ice in upper levels of storms.

In addition to EarthCARE, there are also several planned missions with radar concepts that aim to produce measurements of vertical velocity or surrogates of this vertical motion. NASA's INvestigation of Convective UpdraftS (INCUS), applies a novel sampling strategy using three Ka-band non-Doppler radars in a train to measure the change in reflectivity at short time scales (30, 90 and 120 seconds) to derive the vertical motion and convective mass flux information about storms \citep{Prasanth2023,Dolan2023}. INCUS is currently slated to launch in late 2026 and will be followed with two Doppler radars as part of NASA’s Atmospheric Observing System (AOS).  One radar will operate at Ku-band in an inclined orbit and with a performance similar to the GPM KuPR and a second will operate at higher frequencies (Ka- and W- or both) in a polar orbit. At the time of writing this the second radar is still in its formulation phase.

Past papers have compared the profiles of reflectivity measured by CloudSat, TRMM and GPM. Some papers compared sensors to one another \citep[TRMM to CloudSat;][]{SindhuandBhat2013}, and others compared the profiles with  ground-based radars \citep{Fall2013}, showing the strengths of each radar. CloudSat was designed to observe clouds, the formation of precipitation and as well as precipitation that goes undetected by GPM \citep[e.g.,][]{Berg2010,Behrangi2016}, whereas TRMM and GPM were designed to detect larger hydrometeors and more intense liquid precipitation rates. Beyond the measured reflectivity profiles, retrieved precipitation rates, both rain and snow, have been compared between the satellites and ground-based estimates. Initial snow estimates from GPM were substantially lower than those of CloudSat \citep{Casella2017,Skofronick-Jackson2019,Mroz2021}, but could be improved by adjustments to assumptions in the solid phase retrievals shown by \citet{Chase2020,Chase2021,Chase2022}. For liquid precipitation, \citet{HaydenandLiu2018} suggested the use of CloudSat precipitation retrievals for the lower precipitation rates (approximately 1 $\mathrm{mm \ hr^{-1}}$) and GPM for the larger precipitation rates to help create a more complete picture of the global precipitation. \citet{Liang2024} also suggested a combined dataset between CloudSat and GPM's KaPR to better capture the full profile of hydrometeors.  

Interpretation of radar echoes suffers from complicating factors, including effects of attenuation \citep[e.g.,][]{MeneghiniandKozu1990,Lhermitte1990}, multiple scattering \citep[e.g.,][]{Battaglia2010} and non-Rayleigh scattering that get exacerbated as the operating frequency of the radar increases. These challenges are more prevalent in convection, where vertical motions support the formation of larger hydrometeors. Although these difficulties are understood, there has been a dearth of both qualitative and quantitative comparisons of Ku-, Ka- and W-band reflectivity to assess these effects. 

The research reported in this study compares Ku-, Ka- and W-band radar profiles measured within deep convection. Given Ku-band is the least affected by attenuation, multiple scattering and non-Rayleigh scattering, the primary objective is to compare the Ka-band measured profiles to the Ku-band radar profiles and quantify how much of the Ku-band information content is captured. Ka-band is the primary focus of this evaluation because this paper intends to evaluate how the emerging Ka-band only platforms, such as INCUS and tomorrow.io, will perform in deep convection. Furthermore, the choice of shorter wavelength radars (Ka- and W-) might be the preferred wavelength for future spaceborne missions because of the added cost of Ku-band systems and the difficulties in accommodating the requisite antenna sizes on smaller space platforms  (i.e., Ku-band systems require a larger antenna to achieve a similar sensitivity to Ka-band and W-band). This study further compares the Ku- and Ka-band profiles to those obtained by the W-band CPR of CloudSat. The intent of this comparison is not to highlight the benefits of combining the information provided by these different frequencies, as this will be a topic of a future study, but rather to provide some insight, both qualitative and quantitative, on the attenuation, non-Rayleigh scattering, and multiple scattering experienced by the millimeter wavelength radars (W-, and Ka- band) within deep convection in contrast to the Ku-band reflectivity. The aims of the study are achieved both by analysis of coincident GPM and CloudSat observations complied in the \citet{Turk2021} database and by analysis of the entire Ku- and Ka- band reflectivity GPM record. The primary focus of the analyses is limited to echoes above the melting level and near the tops of intense deep storms where convective updrafts are often maximized \citep[e.g.,][]{Varble2014}.

\section{Data and Methods}

\subsection{Radar descriptions}

The two radars used in this study are the CloudSat Cloud Profiling Radar \citep[CPR; ][]{Stephens2002} and the Global Precipitation Measurement mission’s Dual-frequency Precipitation Radar \citep[DPR; ][]{Hou2014}. The CPR is a W-band (94 GHz; 3 mm) non-scanning radar (i.e., single beam) with a vertical resolution of 500 m, a horizontal resolution of about 1.4 km and a minimum sensitivity of about -30 dBZ early in the life of the radar to approximately -25 dBZ toward the end of the mission life. The CPR was launched in 2006 into a sun synchronous polar orbit with an inclination of 98.2 degrees as part of NASA’s A-train \citep{Stephens2008}. The approximate overpass times of CPR is 1:30 am/pm local time at the equator. While the radar was initially planned to operate for 24 months, it succeeded in operating for 17 years and the mission formally ended in December 2023. In 2011, a spacecraft battery anomaly occurred which prevented CPR from collecting observations during night (i.e., no sunlight on the solar array). CPR collected data regularly in the daylight only mode until the end of the mission, with interruptions to the record occurring because of reaction wheel failures and the introduction of a solution designed to sustain the measurement record \citep[e.g.,][]{Hallowell2022}.

The DPR was launched in early 2014 into a 65-degree inclination operating at Ku- (KuPR; 13 GHz; 2 cm) and Ka-band (KaPR; 35 GHz; 8 mm). The KuPR and KaPR are scanning radars with 49 cross-track beams creating a 250 km swath. The KuPR has a vertical resolution of 250 m, a horizontal resolution of about 5 km, and a minimum sensitivity of near 15 dBZ \citep{Masaki2022}. The KaPR has two scan modes, one that was originally matched to the inner swath of the KuPR, which has a vertical resolution of 250 m and a minimum sensitivity of 19 dBZ. The second scan mode was originally located interlaced in the inner swath of the KuPR, but these scans were moved in May 2018 to match the outerswath of the KuPR enabling full swath coincident dual-frequency measurements \citep{Furukawa2018}. The vertical resolution of the second scan mode is 500 m and has a minimum sensitivity near 14 dBZ. The only notable complications of the GPM bus are that the satellite is down to three reaction wheels and there was an orbit raise in November 2023. Data collected after November 2023 are currently under quality control with the GPM team. 

\subsection{GPM-CloudSat Coincidence}
The coincidence dataset between GPM and CloudSat \citep[2B.CSATGPM; ][]{Turk2021} is quantitatively used to compare profiles available from the three wavelengths of the two radars operating in space. The dataset is made by finding all points within the GPM swath that intersect the CloudSat ground track within 15 mins of one another from 2014 through 2019. The data are collocated by taking the closest footprint of GPM to each footprint of CloudSat. See \citet{Turk2021} for more specific details of the matching procedure. From all coincidences, we sub-sample the dataset to only convectively labeled profiles from the GPM algorithm \citep[i.e., ][]{Awaka2021} and profiles that extend at least 5 km above the surface elevation, where the top was determined from the GPM storm top algorithm (heightStormTop). These criteria were used to isolate deep convection. The result is approximately 5,000 profiles of coincident measurements globally and a map of their locations can be found in the appendix (Fig. S1).

\subsection{GPM-DPR dataset}
To extend the multi-wavelength comparison of storms to a larger sample size, the entire GPM-DPR dataset, ranging from March 2014 to November 2023, is also analyzed. Both the 2A.DPR files and the 2B.CMB files are acquired to enable the use of both the uncorrected measured radar reflectivities (2A.DPR) and the combined radar-radiometer retrievals of precipitation rate and water content \citep{Grecu2016,Olson2022}. Data are sub-sampled in the same way as the coincident dataset. After all subsampling, more than 39 million profiles are available for the analysis described in Section 3.2.

\subsection{Convective Proxies}

\begin{figure}[t]
 \begin{center}
 \includegraphics[width=3in]{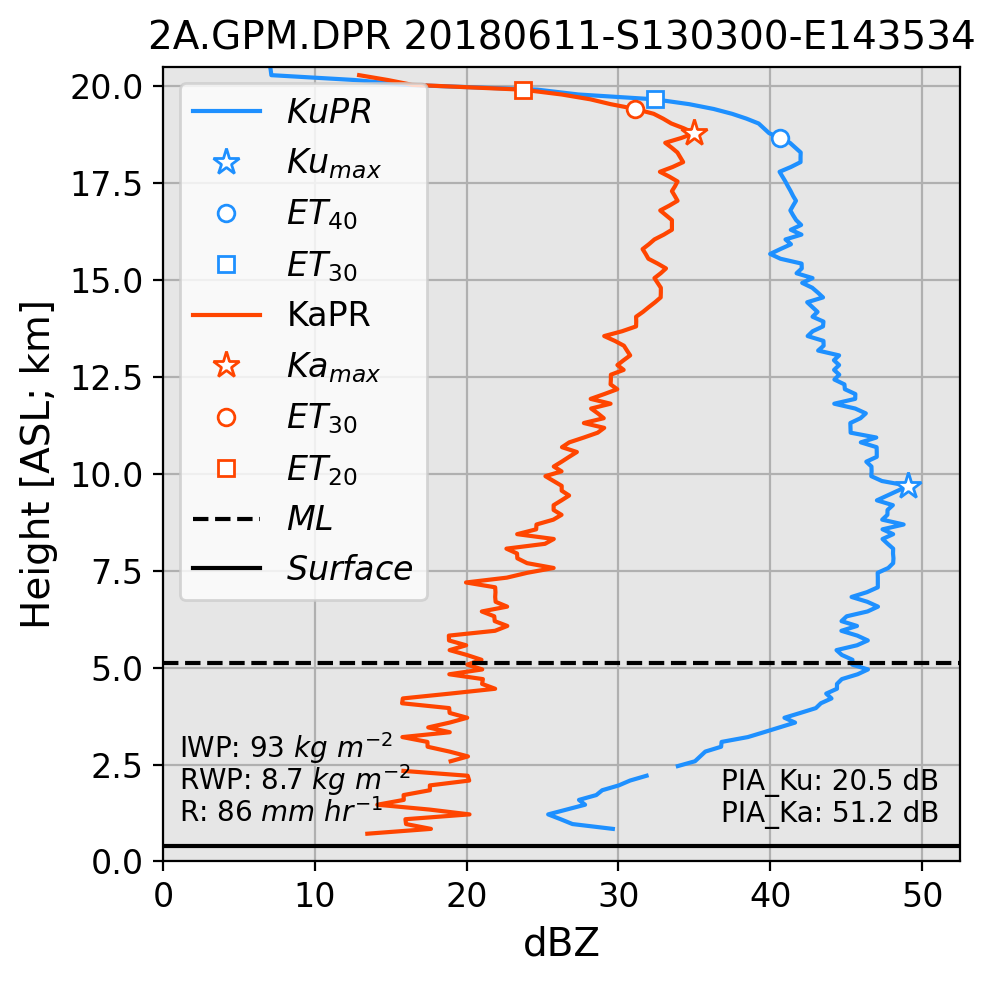}\\
 \caption{Example profile from GPM-DPR over the Arabian Peninsula on 11 June 2018. Several convective proxies are shown including the 20 dBZ (square red marker) and 30 dBZ (circle red marker) echo-top for the KaPR, the maximum of KaPR above the melting level (red star), 30 dBZ (square blue marker) and 40 dBZ (circle blue marker) and the maximum of KuPR above the melting level (blue star). The horizontal black lines are the melting level (dashed) and the surface (solid). The text on the bottom of the figure is the combined retrieval of ice water path (IWP), rainwater path (RWP) near surface precipitation rate (R), two-way path integrated attenuation at Ku and Ka (PIAKu and PIAKa respectively).}\label{f1}
 \end{center}
\end{figure}

A wide variety of convective proxies are used to compare the results here with those in the literature to understand how these proxies differ according to the operating frequency of the radar measurements they are derived for. Echo-tops of 0 dBZ and 10 dBZ are extracted from the CPR profiles, 20 dBZ and 30 dBZ are extracted from the KaPR, and 30 dBZ and 40 dBZ are extracted for the KuPR. We also leverage the heightStormTop product which is the top of the DPR data determined by a coherent precipitation echo in the KuPR data \citep{Iguchi2021}. Beyond the echo-tops we also extract a variety of retrieved quantities from the 2B.CMB retrieval \citep[Version 7; ][]{Olson2022,Grecu2016}. Specifically, we use the water content split into ice water path (IWP) and rain water path (RWP), the near surface instantaneous precipitation rate (R), and the path integrated attenuation (PIA) at Ku- and Ka-band. Lastly, the height of the maximum reflectivity above the melting level (MAML) is extracted. A larger MAML value indicates a stronger convective updraft capable of lofting large hydrometeors to higher altitudes. We use the melting level instead of the surface as the reference point partly to avoid the influence of the radar bright band, which can have large reflectivity values, but also because convective updrafts tend to be strongest above the melting layer \citep{Varble2014}. The melting level is determined from the 2A.DPR algorithm \citep[][]{Iguchi2021}, which uses the bright band location if one is detected, or the 0°C isotherm from a numerical weather prediction model if not. An example of convective profiles with annotated convective proxies can be found in Fig. \ref{f1}. For this case, collected on the west coast of the Arabian Peninsula, the maximum reflectivity from KaPR is found at 18 km ASL and 35 dBZ, while the less attenuated KuPR is lower in the atmosphere, maximizing at about 9 km ASL and almost 50 dBZ.  
For the global maps of convective proxies (Section 3.c.2) we use the standard anomaly which is defined as
\begin{equation}
    z = \frac{(x - \mu)}{\sigma} \label{e1}
\end{equation}
where $x$ is the KuPR or KaPR data observation, $\mu$ is the mean value of $x$ across all KuPR or KaPR observations and $\sigma$ is standard deviation of $x$ across all GPM observations. The standard anomaly is used to evaluate in a relative sense where stronger convection is located globally across multiple metrics.

\section{Results and Discussion}

\subsection{Case Studies: Coincident Dataset}

Before presenting broad statistical comparisons between CloudSat, KaPR and KuPR, a few examples of coincidences in strong convection are analyzed. Strong convection is chosen to highlight a few cases where attenuation and multiple scattering would most likely occur (i.e., worst case scenario). 

\subsubsection{12 June 2018}

\begin{figure*}[t]
 \begin{center}
 \includegraphics[width=6in]{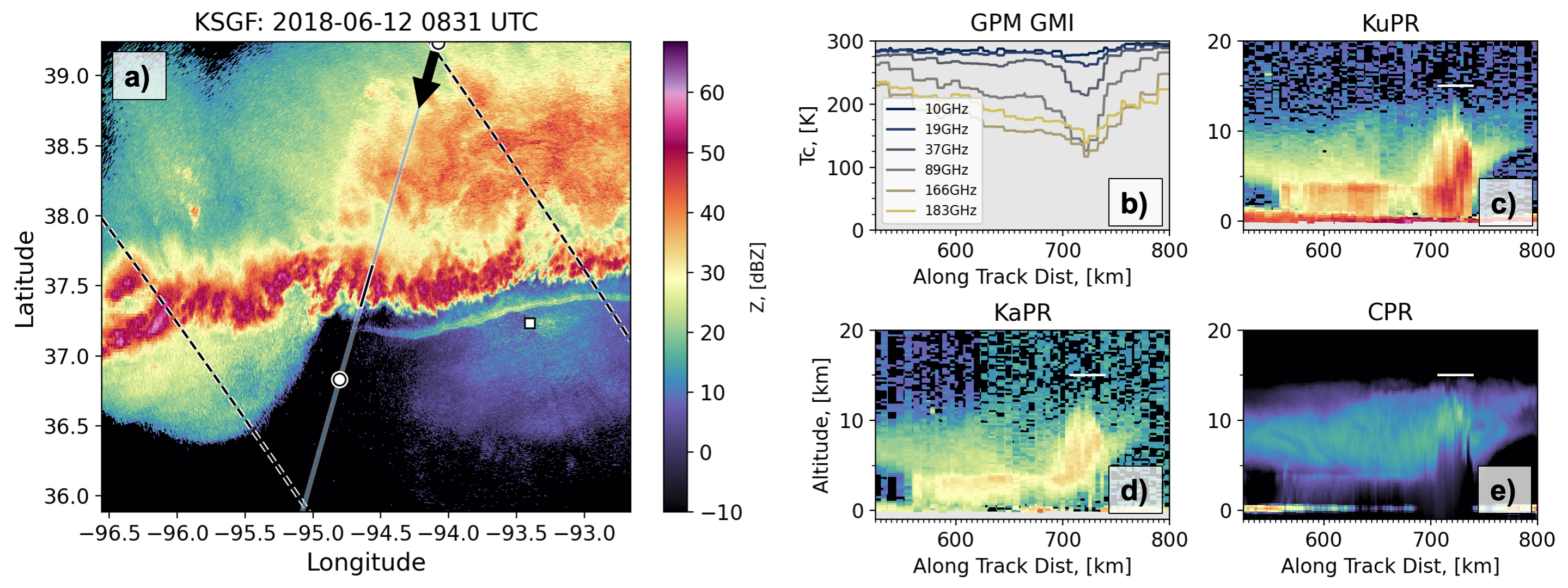}\\
 \caption{A convective coincidence within the GPM-CloudSat dataset exemplifying the view of convection from the three spaceborne wavelengths. For this case, a nocturnal MCS is observed by NEXRAD, GPM and CloudSat at about 0831 UTC on 12 June 2018. (a) plan view of the composite reflectivity from the KSGF radar, located with the square marker. The dashed lines are the GPM swath; the transparent line across the swath is the CloudSat ground-track. The two filled circle markers are the start and end of the cross-section in b-e. (b) Along track calibrated brightness temperatures from GPM-GMI. (c) KuPR along the CPR track. (d) KaPR along the CPR track. (e) CPR reflectivity. All color scales are the same. Horizontal white lines are the location of the mean profile in Fig 3.}\label{f2}
 \end{center}
\end{figure*}

On 12 June 2018 at about 0830 UTC a mesoscale convective system producing surface winds greater than 65 mph and numerous reports of downed trees (SPC storm reports, not shown) was sampled by the ground-based S-band (KSGF), GPM and CloudSat all within 5 minutes of each other near Springfield, Missouri, USA. The plan view of the scene is shown in Fig. \ref{f2}a, where the ground-based radar composite reflectivity exemplifies a prototypical nocturnal MCS with a leading line of convection and trailing stratiform region. GPM's microwave radiometer (GMI) onboard the same spacecraft as DPR provides additional context to the ongoing convection, showing depressions in the brightness temperatures from 19 GHz and higher frequencies. The polarization corrected minimum brightness temperature of 260 K for 19 GHz is right at the cutoff between storms containing hail and not containing hail according to \citet{Mroz2017} (261 K) and is estimated at a 15$\%$ chance of containing hail from \citet{BangandCecil2019}. 

The cross-sections of all frequencies (Fig. \ref{f2}c-e) show the convective echoes associated with the leading line of convection (700 - 750 km on x-axis). All radar frequencies indicate a characteristic bright-band in the trailing stratiform portion of the storm (550 - 700 km). The magnitude in radar reflectivity detected depends on the radar wavelength. The KuPR peaks near 50 dBZ, while KaPR peaks near 35 dBZ and CPR around 20 dBZ. Lastly, a clear attenuation signal is evident in both CPR and KaPR, with reduced surface echoes (700 - 750 km). Even the KuPR has a relatively reduced surface echo near 730 km (Fig. 2c) compared to adjacent scans which suggests some degree of attenuation, although less that at the other two frequencies.

\begin{figure}[t]
 \begin{center}
 \includegraphics[width=2.5in]{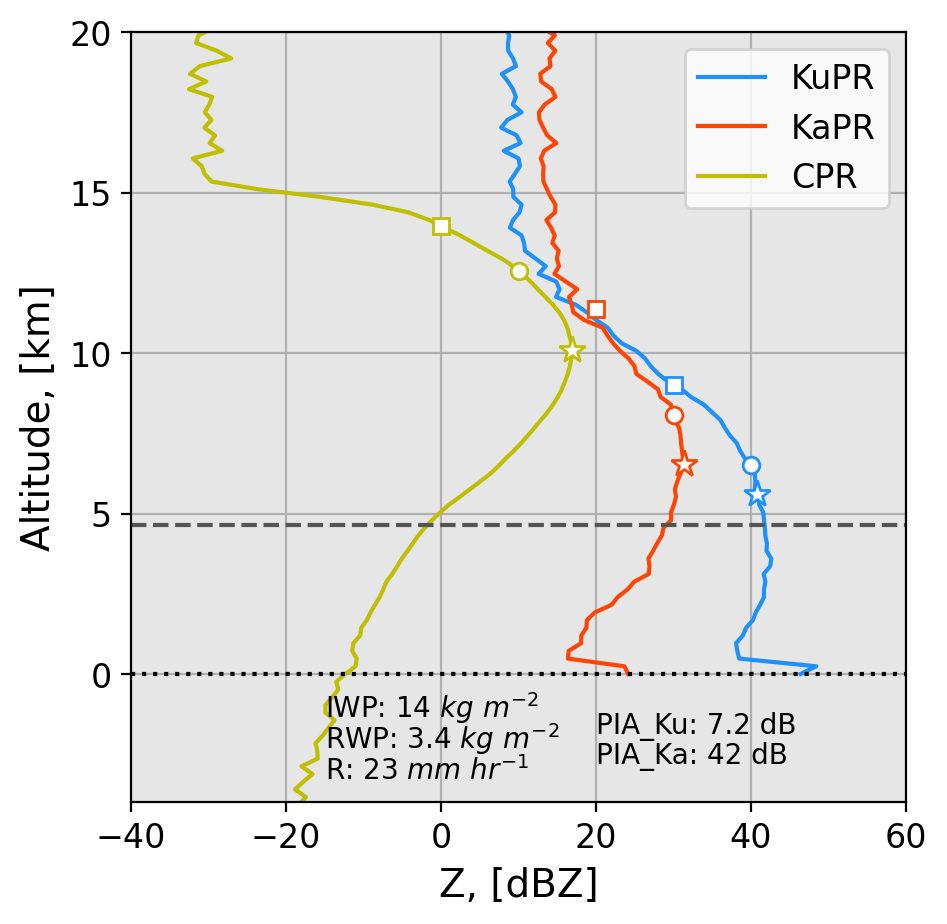}\\
 \caption{Mean reflectivity profiles from along the convective line in Fig 2 b-e showing the different wavelengths of radar for the same location and time. KuPR is in blue, KaPR is in red and CPR is in yellow. Square and circle markers show common echo-top height locations from the literature and the star is the maximum value above the melting level. The dashed black line is the melting level, the dotted line is the surface.}\label{f3}
 \end{center}
\end{figure}

Fig. \ref{f3} presents the mean profile across the leading line of convection (white line in Fig. \ref{f2}b-e). For this case, CPR 0 and 10 dBZ echo-tops are 2 - 4 km taller than the echo-top heights from the KaPR and KuPR. In fact, CPR observes echo up to 15 km, while the KuPR and KaPR only have echo up to 13 and 11 km respectively, primarily because of the lack of sensitivity of GPM’s radars (compared to CPR). The maximum reflectivity for each wavelength occurs at different parts of the profile as well. CPR peaks at 17 dBZ at about 10 km, while KaPR maximizes at 31 dBZ at 6 km, KuPR at 41 dBZ at 5.6 km.

\subsubsection{08 May 2019}

\begin{figure*}[t]
 \begin{center}
 \includegraphics[width=6in]{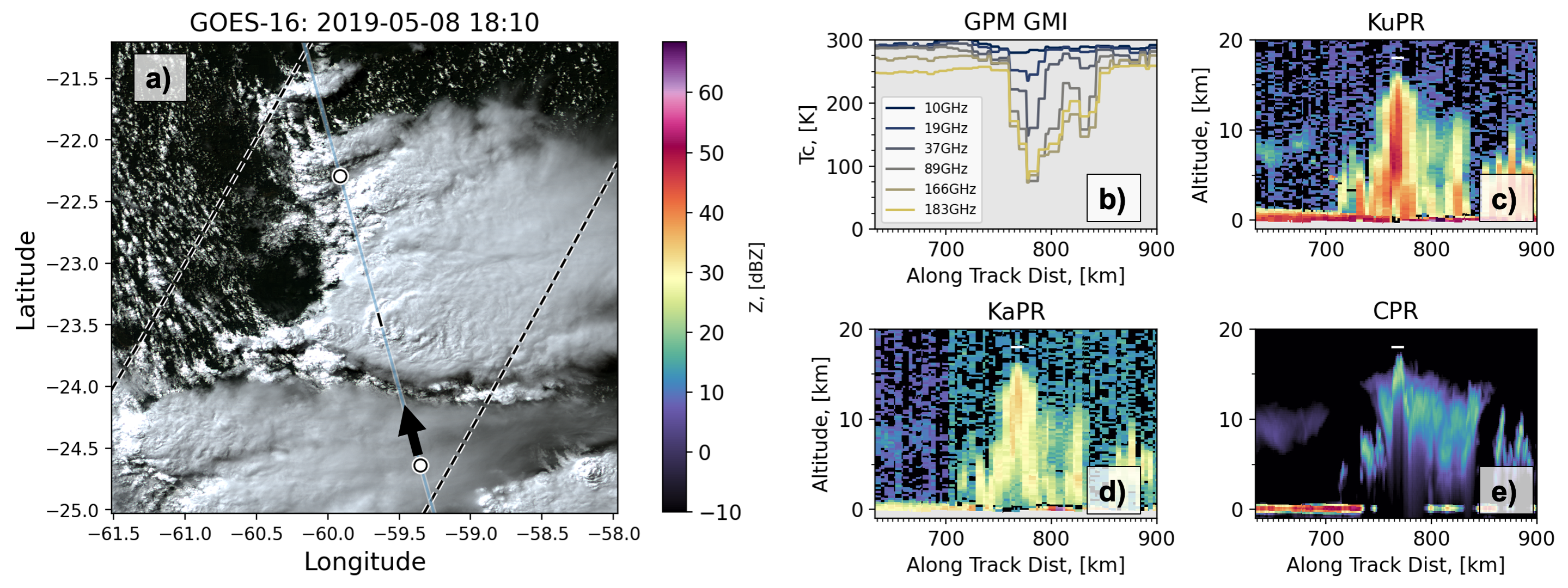}\\
 \caption{A contrasting convective example from the GPM-CloudSat dataset with a more isolated storm morphology. This case is from Paraguay on the afternoon of 08 May 2019 and panel (a) has the GOES-16 visible image instead of the NEXRAD image.}\label{f4}
 \end{center}
\end{figure*}

On 08 May 2019 a storm located over Paraguay was observed by GPM and CloudSat. The GOES visible image from 18:10 UTC shows an overshooting top and above anvil cirrus plume likely indicative of severe weather \citep[Fig. \ref{f4}a;][]{Bedka2018}. Note that the visible image is from 1810 UTC while the CloudSat overpass is from near 1802 UTC and the GPM data are from about 1812 UTC. Despite the about a 10 min difference between GPM and CloudSat a similar structure to the storm is seen with echoes from all wavelengths extending to 15 km, about 5 km taller than the storm sampled in Fig. \ref{f2} (Fig. \ref{f4}c-e). GMI detects polarization corrected brightness temperature depressions from 19 GHz and higher frequencies (Fig. \ref{f4}b). The minimum brightness temperature of 235 K at 19 GHz is indicative of hail \citep[e.g., ][]{Mroz2017} and using the \citet{BangandCecil2019} regression with the 19 and 35 GHz polarization corrected temperatures suggests a $72\%$ likelihood that the column contains hail.

\begin{figure}[t]
 \begin{center}
 \includegraphics[width=2.5in]{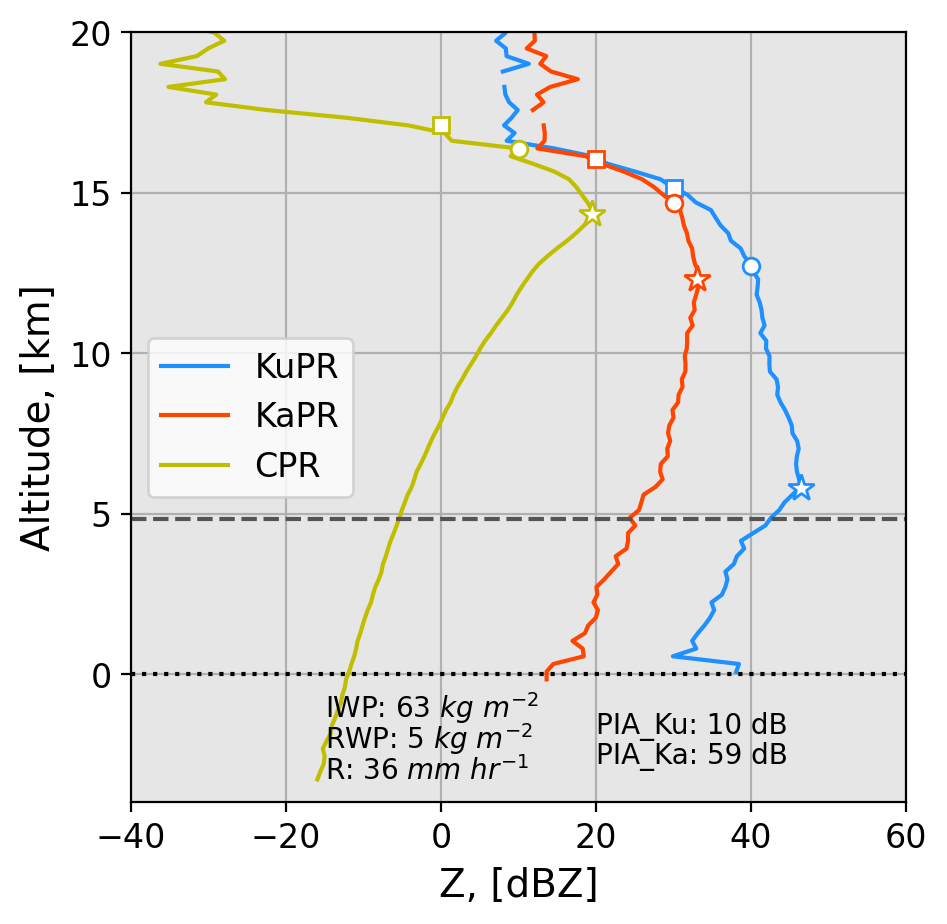}\\
 \caption{As in Fig. \ref{f3}, but derived from the tallest echoes in Fig. \ref{f4} (the Paraguay case)}\label{f5}
 \end{center}
\end{figure}

The mean reflectivity profiles for the core of the storm (tallest echoes) are shown in Fig. \ref{f5} (white line in Fig. \ref{f4}c-e). For this case, all the echo-tops correspond better to one another. CPR’s 0 and 10 dBZ echo-tops are at about 17 and 16 km while KaPR's 20 and 30 dBZ echo-tops are at 15.75 and 13.5 and KuPR's 30 and 40 dBZ echo-tops are at 15 and 12.5 km. The maximum in reflectivity location for each wavelength shows the effects of attenuating radars. CPR and KaPR peak in the upper portion of the column with CPR maximizing at about 20 dBZ around 14 km ASL and KaPR maximizing at 33 dBZ around 12.3 km ASL. As in the previous case study, the maximum of KuPR is found at a lower altitude than for the other frequencies with 47 dBZ at 6km.

\subsubsection{Case Study Discussion}
For both cases, the general structure of the core convection of the storms aloft ($>$ 5km ASL) is captured no matter the wavelength, radar resolution (i.e., along-track spacing) or sensitivity. The differences arise from the echo-top height locations, the anvil characteristics and the depth at which the shorter wavelengths become affected by attenuation and multiple scattering. Contextualizing these two cases of very strong convection in the broader convective proxy literature, both cases would be used in CPR studies of deep convective clouds with 10 dBZ echo-tops exceeding 10 km \citep[e.g., ][]{Takahashi2023}. Meanwhile from the KuPR perspective, the MCS on 12 June 2018 would be within the top $0.33\%$ of storms and the more isolated storm from 08 May 2019 would be within the top $0.03\%$ within the \citet{Zipser2006} framework \citep[see Fig. 6 in ][]{Skofronick-Jackson2018}. In the University of Washington classifications both storms are considered the Deep and Wide classification \citep[][]{Houze2007}, but note the part sampled by the three radars for the MCS case was likely not the most intense part of the storm (stronger GMI brightness temperature deficits were found outside the DPR swath; not shown). 

These two cases provide insight on how to interpret single frequency missions when examining the strongest convective storms. It is known that shorter wavelength radars suffer from more attenuation \citep[e.g., Table 4.7 in ][]{MeneghiniandKozu1990} and multiple scattering effects \citep[e.g., ][]{Battaglia2010}, which suggests missions that only have Ka- or W- band radar are likely to have complications with retrieving information within deep convection. Both examples discussed were intense convection and contained hydrometeors responsible for strong attenuation and multiple scattering. The combined radar-radiometer retrieval from GPM shows large amounts of ice mass and estimated the amount of two-way path integrated attenuation at 42 and 59 dB for Ka-band for the two cases respectively. The goal of this paper is to assess the information contained in the signals of shorter wavelength radars in deep convection, and evaluate the associated attenuation, multiple scattering, and non-Rayleigh scattering. The cases presented here demonstrate all of these challenges and are within tail of the distribution of deep convective samples. Despite the strong nature of these convective profiles, the Ka-band information matches well with the Ku-band for about 5 km of depth (i.e., echo-top to 6 km in the first case). This is encouraging for Ka-band only missions, like INCUS and tomorrow.io, because there is clear signal from the Ka-band in the upper portions of the storms where convective motions, and thus convective mass flux, are the greatest \citep[e.g., ][]{Varble2014}. The W-band has a similar signal but, as expected, the depth of profile where it mimics the Ku-band is limited to about the first 2 km (starting from the KuPR echo-top). In the next section, we use the entire coincident dataset to draw some statistical relationships between echo-top heights between sensors as well as the general profile characteristics.

\subsection{Coincident Dataset Statistics}

\begin{figure}[t]
 \begin{center}
 \includegraphics[width=2.5in]{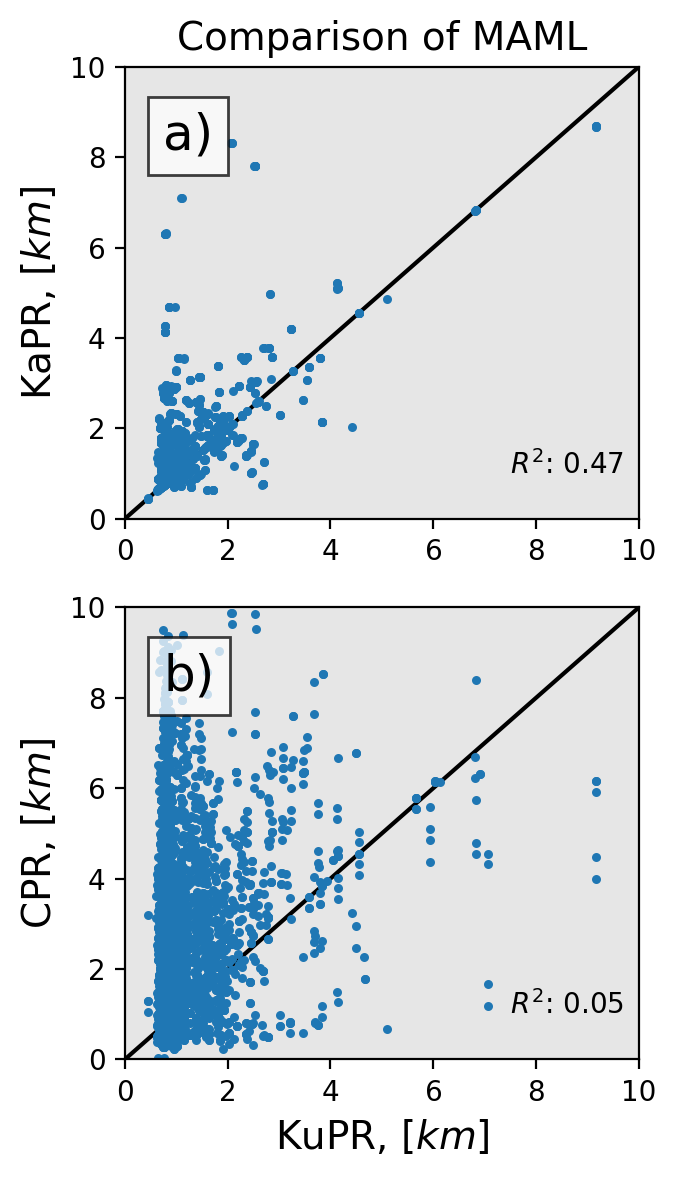}\\
 \caption{The potential relationship between the height of the maximum reflectivity above the melting level (MAML) between (a) KaPR and KuPR and (b) CPR and KuPR. }\label{f6}
\end{center}
\end{figure}

Using the full coincident dataset (approximately 5000 points) we first compare the height of the maximum of reflectivity above the melting level (MAML) between the shorter wavelengths (Ka- and W- band) and Ku-band (Fig. \ref{f6}). It is not expected that the MAML for the shorter wavelengths should be identical to Ku-band because of the nuances of attenuation and non-Rayleigh scattering but the magnitude of the correlation could show that the shorter wavelengths have a similar signal in convective intensity. For the entire database considered, the MAML derived from Ka-band compared to Ku-band shows a weak linear relationship ($R^2$: 0.47; Fig. \ref{f6}a) while W-band and Ku-band have no linear relationship ($R^2$: 0.05; Fig. \ref{f6}b). 

\begin{figure}[t]
 \begin{center}
 \includegraphics[width=2.5in]{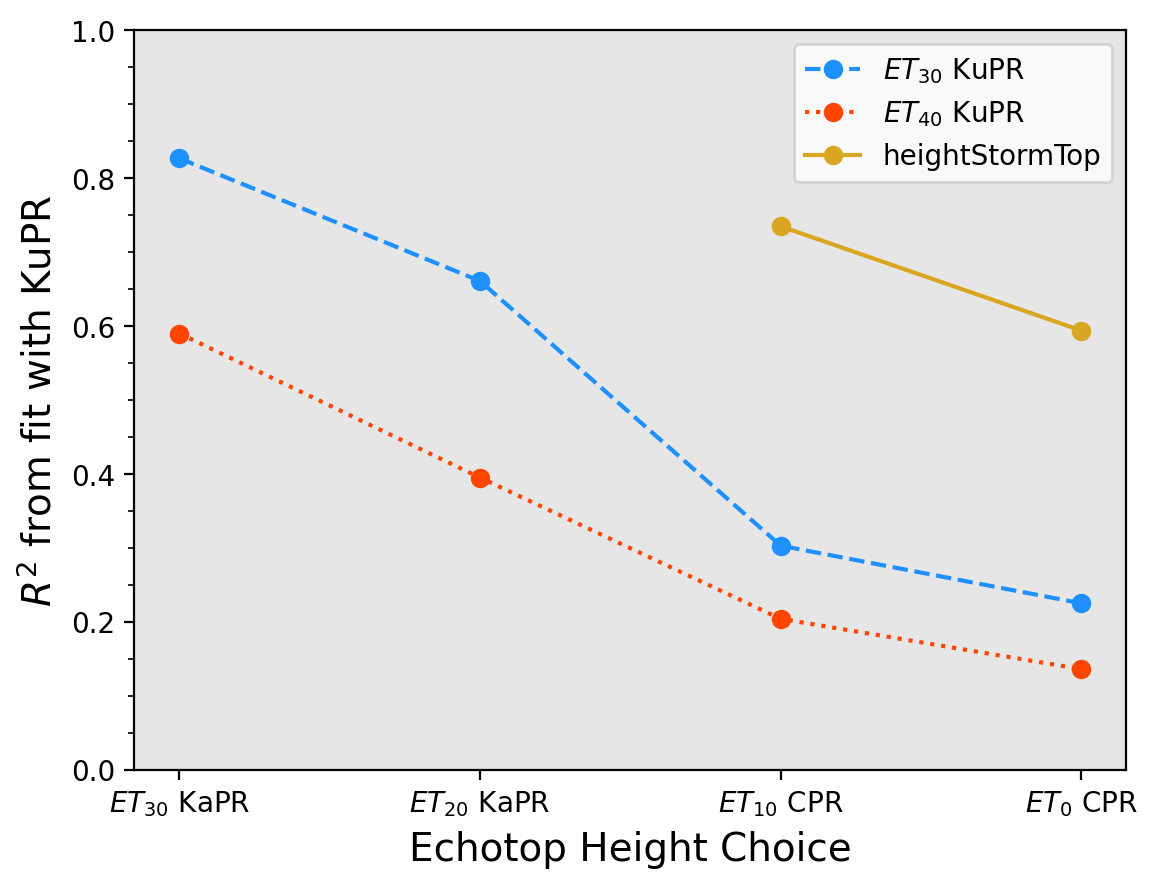}\\
 \caption{The strength of a linear relationship (coefficient of determination) between radars for different echo top height thresholds in the combined GPM-CloudSat dataset.}\label{f7}
 \end{center}
\end{figure}

The correspondences between the radar profiles are measurably closer when expressed in terms of echo top heights (Fig. \ref{f7}). $R^2$ values follow expectation with generally decreasing $R^{2}$ with larger differences in reflectivity thresholds. The largest $R^2$ values are between the 30 dBZ KaPR height and the 30 dBZ KuPR height, with a value of 0.83 (blue line Fig. \ref{f7}), while the smallest $R^2$ is between CPR’s 0 dBZ echo-top height and KuPR's 40 dBZ echo-top height with a value less than 0.15. Considering the top of the KuPR echo (heightStormTop), the CPR 10 dBZ echo-top corresponds reasonably well with an $R^2$ of 0.74. All linear fits between different echo-tops are shown in Table 1, along with the root mean squared error if the linear fit was used to translate KaPR or CPR echo-top height to a KuPR echo-top height. For example, if the KaPR 20 dBZ echo-top was translated to a KuPR 20dBZ echo-top, the root mean squared error is 0.77 km. Note that the linear regression coefficients in Table 1 for the KaPR to KuPR echo-tops are near 1 for both the 20 dBZ and 30 dBZ echtops, implying you can use the KaPR echo-top interchangeably with the KuPR echo-top. 

\begin{figure*}[t]
 \begin{center}
 \includegraphics[width=6in]{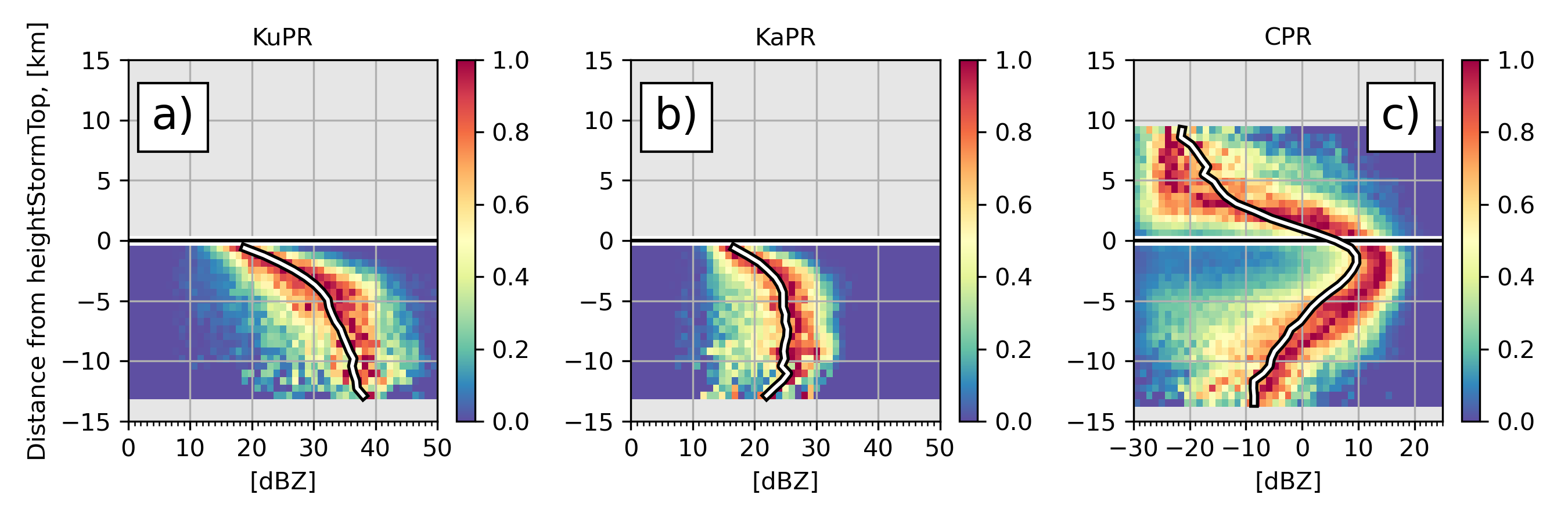}\\
 \caption{Broad statistical profile characteristics from the combined dataset. These are contour frequency by altitude diagrams for (a) KuPR (b) KaPR and (c) CPR for the coincident dataset relative to GPM’s heightStormTop. The color shading is relative frequency normalized to the maximum in each y-axis row. The white line is the median value for each y-axis row.}\label{f8}
 \end{center}
\end{figure*}

Contoured frequency by altitude diagrams (CFADS) oriented from the heightStormTop (i.e., KuPR top; approximately 12 dBZ) characterize the broad shape of the profiles between the three frequencies for the entire database. For KaPR, the median (line Fig. \ref{f8}a) of the CFAD increases at 3 dB/km toward about 4 km where the slope changes to 1 dB/km. Meanwhile KaPR (Fig. \ref{f8}b) increases at a rate of 2 dB/km to 3 km from GPM echo-top and stays constant at about 25 dBZ toward lower altitudes. Lastly, CPR shows echoes deep above the GPM echo-top, sometimes more than 5 km taller (Fig. \ref{f8}c). The slope in the W-band remains constant from about 2.5 km above the GPM echo-top to about 1 km below the GPM echo-top at 5.7 dB/km and decreases at a constant rate of 1.7 dB/km towards lower altitudes.

The coincident dataset gives both a qualitative and quantitative depiction of deep convection from the perspective of radar reflectivity at the three wavelengths currently available from space. In general, the coincident data results suggest that the maximum in both Ka-band and W-band does not correspond well to the maximum in the Ku-band. This is expected given the different sensitivities of each radar and the added challenges of non-Rayleigh scattering, attenuation, and multiple scattering in interpreting the reflectivities. That being said, the echo-top height of the Ka-band 30 dBZ corresponds well to the Ku-band 30 and 40 dBZ echo-top heights. There is then potential to translate between the previous Ku-band only literature to the new Ka-band only literature of future missions (i.e., INCUS; Tomorrow.io).

\begin{table*}
\begin{center}
\begin{tabular}{ |p{3cm}||p{0.5cm}||p{1.5cm}||p{1cm}||p{1.5cm}|}
\hline
\multicolumn{5}{|c|}{Linear Regression Stats: Coincident Dataset} \\
\hline
Pairing & $R^{2}$ & coefficient & intercept & rmse [km] \\
\hline
$\mathrm{KaPR_{MAML}}$ - $\mathrm{KuPR_{MAML}}$ & 0.47 & 0.93 & 0.37 & 0.78 \\
\hline
$\mathrm{KaPR_{20}}$ - $\mathrm{KuPR_{20}}$ & 0.89 & 0.90 & 1.12 & 0.77 \\
\hline
$\mathrm{KaPR_{30}}$ - $\mathrm{KuPR_{30}}$ & 0.83 & 0.95 & 1.27 & 1.05 \\
\hline
$\mathrm{KaPR_{30}}$ - $\mathrm{KuPR_{40}}$    & 0.59 & 0.60 & 0.77 & 1.31 \\
\hline
$\mathrm{CPR_{MAML}}$ - $\mathrm{KuPR_{MAML}}$ & 0.05  & 0.54 & 1.96 & 1.83 \\
\hline
$\mathrm{CPR_{0}}$ - heightStormTop & 0.51  & 0.55 & 3.28 & 1.67 \\
\hline
$\mathrm{CPR_{10}}$ - heightStormTop & 0.59  & 0.65 & 3.36 & 1.56 \\
\hline
\end{tabular}
\end{center}
\caption{\label{table-1} Linear statistics between various radars for the coincident dataset.}\label{t1}
\end{table*}

\subsection{Full GPM record}

\subsubsection{All profile statistics}

\begin{figure}[t]
 \begin{center}
 \includegraphics[width=2.5in]{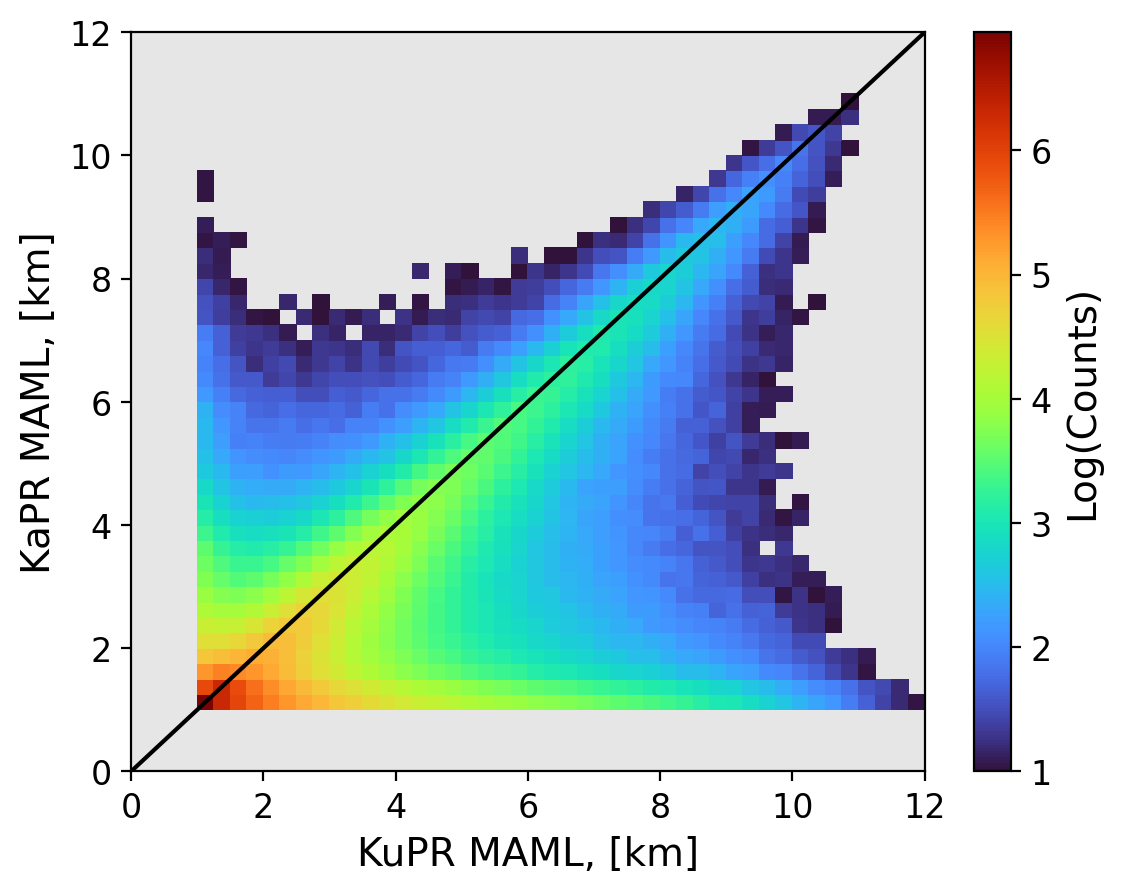}\\
 \caption{The comparison of the MAML of KuPR and KaPR for the entire GPM database.}\label{f9}
 \end{center}
\end{figure}

The number of samples in the coincident GPM-CloudSat dataset is limited. Here we also consider the full GPM record where the KuPR and KaPR are coincident. There are more than 39 million coincident profiles between KuPR and KaPR in deep convection (see Fig. \ref{f11}a) for the distribution of samples). The same analysis presented in Fig. \ref{f6} is reproduced for the larger dataset (Fig. \ref{f9}).

The results of the analysis of the coincident dataset are essentially reproduced in the analysis of the full GPM dataset which is summarized in Table 2. The correspondences between the Ku and Ka band height proxies are closer. For the MAML the $R^{2}$ values of 0.4 suggesting a weak linear relationship between the KaPR and KuPR MAML. That being said, the root mean squared error is less than one kilometer for the linear regression fit on KaPR (0.71 km; Table 2). For brevity, Fig. \ref{f7} is not reproduced for the full GPM dataset but the statistics are included in Table 2. Overall, the various KuPR and KaPR echo-tops correspond well. The $R^2$ values are all greater than 0.6, with the KaPR 30 dBZ echo-top height being able to explain 62$\%$ of the variation of the KuPR 40 dBZ echo-top height. The best linear correspondence ($R^2$ of 0.9) is the 20 dBZ echo-top for both KuPR and KaPR which is anticipated because 20 dBZ should be close to the Rayleigh scattering regime for KuPR and KaPR. This bulk analysis shows that KaPR echo-tops linearly relate to KuPR echo-tops on a global average.  This implies an ability to translate Ka-band echo-tops to Ku-band echo-tops and suggests that Ka-band only missions can provide a similar measure of convective intensity as a Ku-band only mission. It is not clear if this global relationship holds for all locations, for example regions of locally stronger (weaker) convection.

\begin{figure*}[t]
 \begin{center}
 \includegraphics[width=6in]{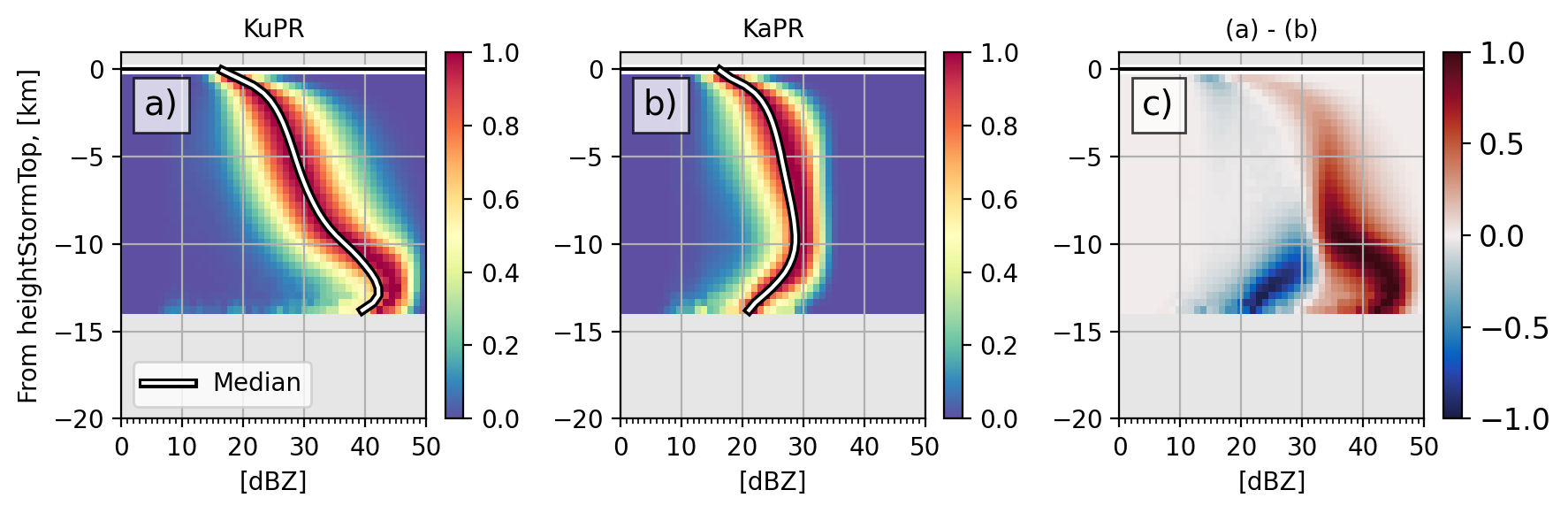}\\
 \caption{A statistical comparison (CFAD) of the top 1$\%$ tallest storms (greater than 15 km above the surface) in the entire GPM record. The profiles are oriented from heightStormTop and normalized to the max count for each y-axis bin. (a) KuPR profiles (b) KaPR profiles. The total number of profiles in these CFADs is 372,249.}\label{f10}
 \end{center}
\end{figure*}

Beyond looking at the echo-tops, we consider the full reflectivity profiles of the KuPR and KaPR. In this case, only that portion of the profiles corresponding to temperatures less than 0degC and profiles where the echo-top is more than 15 km above the surface were considered. These profiles were chosen anticipating that they are profiles with the largest likelihood of attenuation, multiple scattering and non-Rayleigh scattering. Fig. \ref{f10}ab shows that even in the tallest profiles (top 1$\%$ of profiles), the median KaPR profile increases alongside the KuPR profile from GPM defined echo-top (heightStormTop) to 10 km below, maximizing with a value near 30 dBZ. The primary differences in the CFADs are the shift to larger reflectivities at heights between 0 and 10 km below echotop, the more frequent smaller reflectivities deeper than 10 km into the echo, and the slopes of the reflectivity. The shift to larger reflectivities between Ku- and Ka-band (Fig. \ref{f10}c) can primarily be explained by non-Rayleigh scattering effects, leading to reduced radar backscatter cross-sections for Ka-band. The more frequent weaker reflectivities at Ka-band greater than 10 km into the profile are likely from attenuation. As for the differences in the reflectivity slopes,  the KuPR slope near the GPM echo-top is about 5 dB/km for the first km, but then reduces to about 2 dB/km until about 10 km below GPM echo-top where the slope increases again until 12 km. The KaPR CFAD shows a similar 5 dB/km slope for the first km, but then has a reduced slope of 0.5 dB/km till about 10 km and the slope then becomes negative. The CFADs of the tallest 1$\%$ of profiles observed by GPM suggest that despite the added complications of multiple scattering, attenuation and non-Rayleigh scattering at Ka-band, there is similar information contained in the Ka-band signal compared to that of the Ku-band. Thus, Ka-band missions looking to quantify convective intensity, especially missions looking at the ice phase layer (e.g., INCUS), can expect to have usable signal. 

\subsubsection{Global Distributions}

\begin{figure}[t]
 \begin{center}
 \includegraphics[width=3in]{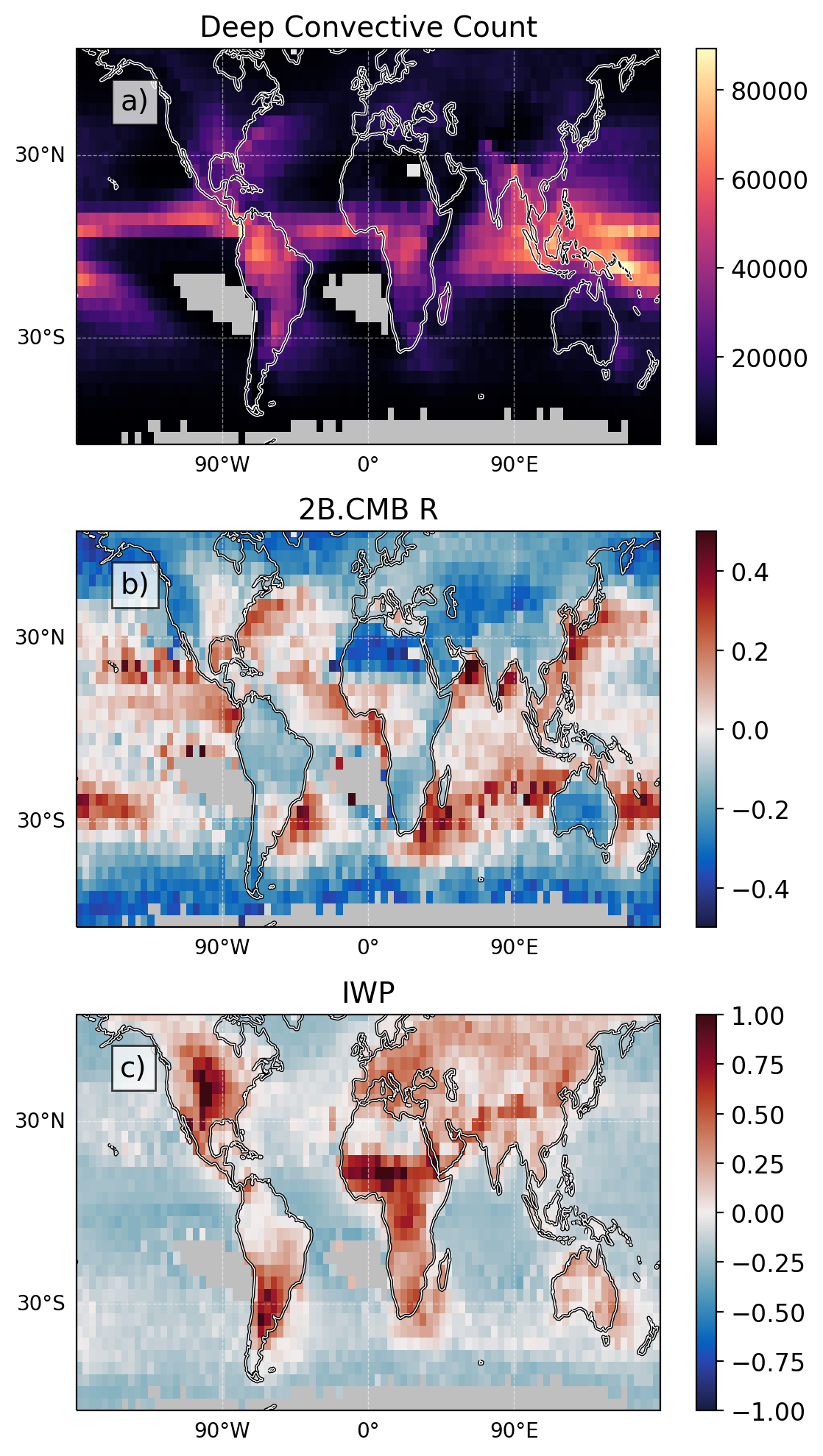}\\
 \caption{Global maps of deep convection as observed by GPM retrieved properties. (a) The count of all deep convective profiles from 2014 – 2023. (b) The conditional mean standardized anomaly of the combined retrieval near surface rain rate. (c) as in b but for the retrieved ice water path.}\label{f11}
 \end{center}
\end{figure}

Before we consider the spatial distribution of the radar echo-tops, we first consider the retrieved quantities out of the 2B.CMB algorithm \citep{Grecu2016,Olson2022} that provide the best estimate of the hydrometer profiles (i.e., mass, precipitation rate) using both KuPR and KaPR as well as the GMI brightness temperatures. Fig. \ref{f11} shows the mean standard anomaly in each latitude-longitude bin.  The original unscaled data of Fig. \ref{f11} and \ref{f12} can be found in the supplementary material (Fig. S2, S3 respectively). 

Using only the instantaneous near surface instantaneous rainfall rate (R) to characterize the locations of the strongest convection, the data suggest the primarily oceanic regions with many hotspots located near coastlines around 30$^\circ$ N and S (Fig. \ref{f11}b). Meanwhile, using IWP as a metric of convective intensity suggests the opposite of rainfall, highlighting mostly continental locations, with hot spots (greater than 1 standard deviation) being found over the central US, central Argentina and central Africa (Fig. \ref{f11}c). Note that these are conditional statistics on deep convection (i.e., when deep convection occurs, these are the global hotspots). The seemingly contradicting results between IWP and R has been shown before, suggesting that the most intense rainfall rate events are associated with weaker convection \citep[e.g., ][]{Hamada2015, Gingrey2018, Zipser2021, Xu2022}. These global distributions will serve as a reference for the various echo-top values examined next.

\begin{table*}
\begin{center}
\begin{tabular}{ |p{3cm}||p{0.5cm}||p{1.5cm}||p{1cm}||p{1.5cm}|}
\hline
\multicolumn{5}{|c|}{Linear Regression Stats: Full GPM record} \\
\hline
Pairing & $R^{2}$ & coefficient & intercept & rmse [km] \\
\hline
$\mathrm{KaPR_{MAML}}$ - $\mathrm{KuPR_{MAML}}$ & 0.41 & 0.93 & 0.35 & 0.71 \\
\hline
$\mathrm{KaPR_{20}}$ - $\mathrm{KuPR_{20}}$    & 0.90 & 0.89 &  1.18 & 0.73 \\
\hline
$\mathrm{KaPR_{30}}$ - $\mathrm{KuPR_{30}}$   & 0.84 & 0.95 & 1.19 & 0.83 \\
\hline
$\mathrm{KaPR_{30}}$ - $\mathrm{KuPR_{40}}$    & 0.62 & 0.61 & 0.91 & 0.98 \\
\hline
\end{tabular}
\end{center}
\caption{\label{table-2} Linear statistics between various radars for the full GPM record.}
\end{table*}

Qualitatively, all echo-top proxies generally highlight the same regions of high IWP: central Africa, central Argentina and central United States (Fig. \ref{f12} a-f). The differences between the KaPR maps and KuPR maps are found in Fig. \ref{f12} g-i. The MAML shows the largest difference over central Africa, with KaPR maxima height more than 0.25 standard deviations higher than that of the Ku-band (Fig. \ref{f12}g). Comparatively, regions of large IWP over the USA and central Argentina show smaller differences in the MAML between KuPR and KaPR. This implies a difference in the structure of convective profiles of KuPR over central Africa, where the MAML is lower in the atmosphere (i.e., closer to the melting level). We hypothesize this is a result of a characteristically different convective environments over central Africa compared to USA or central Argentina, but this is beyond the focus of this paper. However, this serves to highlight how the differences themselves offer hints about the vertical structure of the convection, a topic certainly worthy of future study and an area of interest to INCUS. As for the differences between the KaPR 20 dBZ and KuPR 30 dBZ echo-tops (Fig. \ref{f12}h), in general the high latitude oceans have greater KaPR anomalies where deep convection is far less frequent (Fig. \ref{f11}a) and continental regions show higher KuPR anomalies. Meanwhile the comparison of the 30 dBZ KaPR to the 40 dBZ KuPR shows that the tropical oceanic regions have generally higher Ka-band echo-top anomalies (Fig. \ref{f12}i). Thus, if one were to use a Ka-band only radar system and the 30 dBZ echo-top, there might be a systematic bias in characterizing stronger convection over the tropical oceans compared to Ku-band systems.  In summary, the spatial correspondence between KaPR and KuPR echo-tops match well qualitatively. 

\begin{sidewaysfigure*}[ht]
    \begin{center}
    \includegraphics[width=\textwidth]{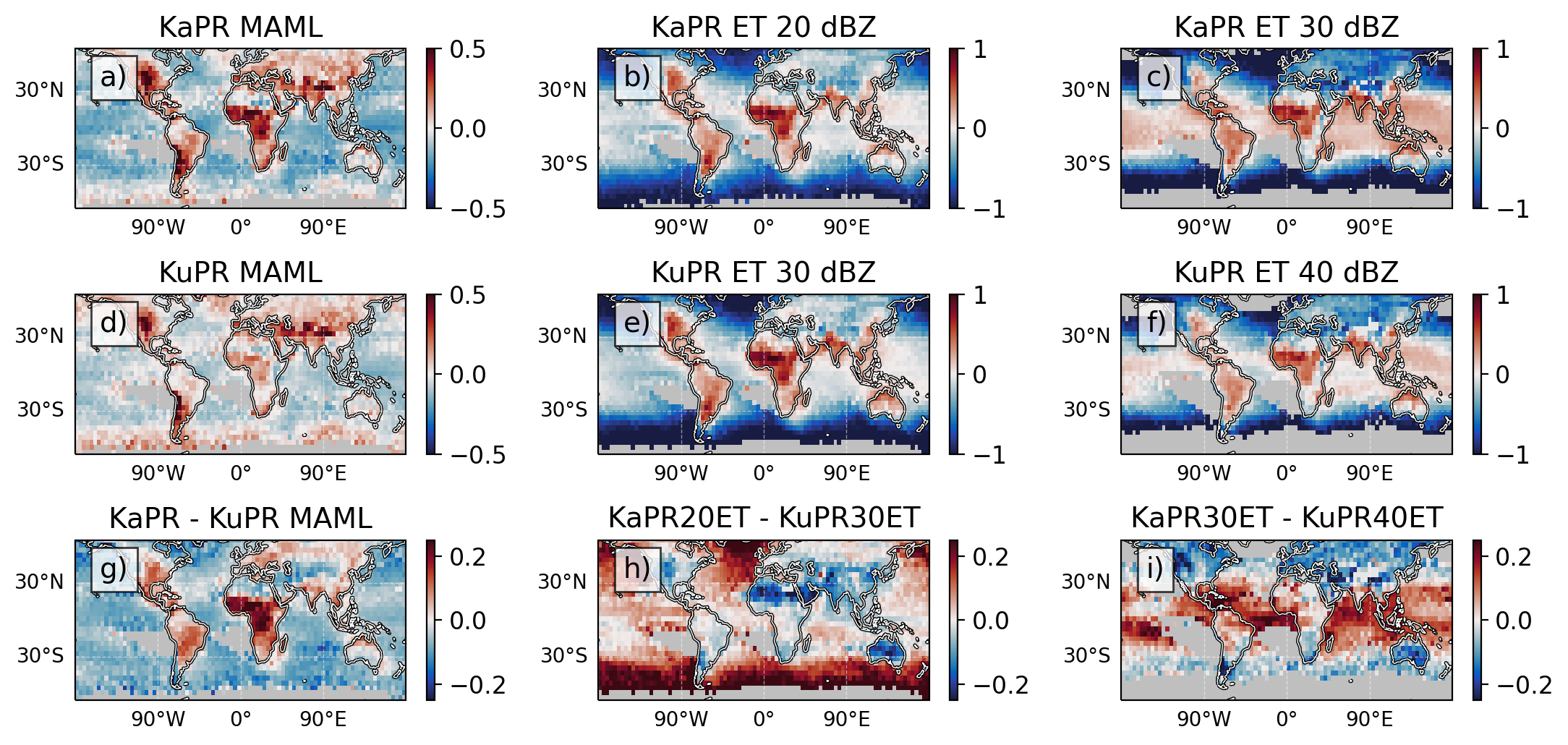}
    \caption{Comparing the spatial distribution of the various measured echo-tops from KuPR and KaPR. (a) The mean standard anomaly of the height of maximum of KaPR above the melting level (b) The mean standard anomaly for KaPR 20 dBZ echo-top (c) The mean standard anomaly for KaPR 30 dBZ echo-top (d) The mean standard anomaly for height of maximum KuPR above the melting level (e) The mean standard anomaly for  KuPR 30dBZ echo-top (f) The mean standard anomaly for KuPR 40 dBZ echo-top (g) difference between panels a and d (h) difference between panels b and e (i) difference between panels c and f. }
    \label{f12}
    \end{center}
\end{sidewaysfigure*}

\section{Summary and Conclusions}

There has been a rich history of spaceborne weather radars going back to TRMM in 1997 and including three main frequencies: W-band (3 mm; CloudSat; CPR), Ka-band (8 mm; GPM KaPR) and Ku-band (2 cm; TRMM and GPM KuPR). Numerous studies \citep[e.g.,][]{Zipser2006,Houze2015,TakahashiandLuo2014} have used the spaceborne measurements to investigate the distribution of global storm properties, but often use only one frequency in their description of storms. Thus, those studies are isolated to their specific frequency and the process of translating the results between frequencies is not trivial. This paper uses coincidences between GPM and CloudSat \citep{Turk2021} as well as the full GPM record to compare the selected convective proxies (i.e., various echo-tops) between radars. The results and conclusions are as follows: 

\begin{enumerate}
\item Two cases of strong convection were analyzed using coincident observations from all three spaceborne radar frequencies:  a prototypical nocturnal MCS over the central USA and a more isolated system over Paraguay. The cases were chosen to exemplify scenarios of strong attenuation, likely multiple scattering and non-Rayleigh scattering effects (i.e., worst case scenarios). Qualitatively, the two cases showed the potential richness of coincident multi-frequency radar information as well as the complicating effects for millimeter wave radars deep convection: (Fig. \ref{f2} - Fig. \ref{f5}).

\item The entire coincident dataset corroborated the qualitative trends observed for the case studies, including the following:  
    \begin{itemize}
    \item Cloudsat’s Cloud Profiling Radar (CPR) tended to maximize near the GPM detected storm top (i.e., heightStormTop; 12 dBZ at Ku-band) and did not linearly correspond to the height of the Ku-band (KuPR) maximum reflectivity ($R^2$ near 0).
    \item GPM’s Ka-band (KaPR) reflectivity maximized at lower altitudes than the CPR, often 4-5 km from GPM heightStormTop, but does weakly linearly relate to the height of the KuPR maximum in reflectivity ($R^2$ of 0.47).
    \item Specific KaPR echo-top values produced reasonable linear fits to specific KuPR echo-top values (e.g., 30 dBZ from KaPR to 40 dBZ from KuPR; $R^2 of 0.59$). 
    \end{itemize}

\item Extending the KaPR and KuPR comparisons to the full GPM record (April 2014 - November 2023; more than 39 million profiles) showed the following:
    \begin{itemize}
    \item The full GPM record had a similar result to the coincident dataset where the linear relationship between the height of the KaPR maximum reflectivity and the height of the Ku-band maximum reflectivity was not strong ($R^2$ of 0.4).
    \item The KaPR 30 dBZ echo-top linear relationship was reasonably strong ($R^2$ of 0.62) with the KuPR 40 dBZ echo-top and had a root mean squared error less than one kilometer. 
    \item CFADS in the tallest 1$\%$ of profiles show that the median KaPR profile follows the KuPR profile with increasing reflectivity till 10 km below GPM heightStormTop.
    \item Maps of echo-tops from KaPR highlight similar regions as KuPR and coincide well with the areas of large ice water path. 
    \end{itemize}
\end{enumerate}

Note that the conclusions from the coincident dataset between GPM and CloudSat are from a relatively small sample size (5000 samples). Unfortunately, several issues with the CPR spacecraft reduce the total size of the coincident dataset, namely the battery failure in 2011 and the loss of most of the reaction wheels to accurately point the radar in 2020. A similar W-band Doppler radar on EarthCARE, launched in May 2024, offers an opportunity to expand such coincidences between GPM and EarthCARE. Furthermore, the EarthCARE Doppler radar information could be leveraged in new ways. Another caveat of this work is that the primary vehicle of analyses were linear relationships (i.e., linear regression) which were used for the sake of simplicity and brevity. Non-linear methods, like random forests or neural networks, have shown good performance in the atmospheric sciences \citep[e.g.,][ and references therin]{Chase2022,Chase2023} and might provide more accurate translations from W-band or Ka-band to Ku-band. Lastly a complicating factor to the interpretation of the work presented here is Non-Uniform Beam Filling (NUBF). \citet{Mroz2018} showed that hail cores never fill the entire GPM-DPR field of view. Thus, when hail is present, which is often true in deep convective storms, NUBF is present, complicating the interpretation of the reflectivities and the subsequent retrievals. 

Despite these limitations, the statistical similarity between the KaPR and KuPR and the potential linear fits between KaPR echo-top height and KuPR echo-top height in this paper supports the assertion that Ka-band only missions (e.g., INCUS) will be able to provide an important characterization of the global distribution of convection. Furthermore, through the comparison of CFADs of the top $1\%$ tallest storms, the KaPR provides valuable insights into even the most intense convective storms. This said, this paper primarily focused on the region of deep convective profiles above the melting level which is the likely location of the strongest convective updrafts \citep[e.g.,][]{Varble2014}. The use of a Ka-band only system for rainfall mapping will have the added challenge of strong attenuation with liquid phase hydrometeors, but these issues could potentially be mitigated with incorporation of other constraints \citep[e.g.,][]{LEcuyer2002} such as coincident radiometric measurements from microwave radiometers or opportunistic passive microwave temperatures from the radar itself \citep[e.g., ][]{Battaglia2020a}.

In an ideal scenario, future spaceborne radar missions looking to characterize convective clouds would have all three available wavelengths to properly measure the wide spectrum of convective processes and associated hydrometeor properties and provide multiple constraints for retrievals. Furthermore, ideally at least one of the available wavelengths would have Doppler capabilities. Unfortunately, such a system is likely not feasible for at least the next decade, but as the cost of spaceborne radars (both launching and building) continues to decrease, agencies should continue to explore multifrequency radar systems.

\clearpage
\acknowledgments
This work was supported by INCUS, a NASA Earth Venture Mission, funded by NASA’s Science Mission Directorate and managed through the Earth System Science Pathfinder Program Office under contract number 80LARC22DA011. K. Rasmussen was also supported by NASA Grant number 80NSSC22K0608. We would like to thank Sarah Bang, Pavlos Kollias, Zhoucan Xu and Stephen Nesbitt for their input to this manuscript.

%
%
\datastatement
The GPM DPR dataset used in this paper can be found with this doi: 10.5067/GPM/DPR/GPM/2A/07, while the combined retrieval data can be found with this doi: 10.5067/GPM/DPRGMI/CMB/2B/07. The coincident dataset doesn’t have a DOI, but can be found on \url{https://storm.pps.eosdis.nasa.gov/storm/} or within the arthurhou data system (\url{https://arthurhouhttps.pps.eosdis.nasa.gov/gpmdata/}). The processing scripts used in this paper and the processed datasets will be available on Dryad upon publication. 

%






%



\bibliographystyle{ametsocV6}
\bibliography{references}

\end{document}